\documentclass[11pt]{article}
\usepackage{graphics}

\newcommand{\nn}{\nonumber}
\newcommand{\lesssim}{\stackrel{<}{ _{\sim}}}
\newcommand{\grsim}{\stackrel{>}{ _{\sim}}}

\begin{document}

\title{On the tidally induced gravitational collapse of a particle cluster}
\author{Kashif Alvi\footnote{Theoretical Astrophysics 130-33,
California Institute of Technology,
Pasadena, California 91125} \and Yuk Tung Liu\footnotemark[1]}
\date{}
\maketitle

\begin{abstract}
An important issue in the dynamics of neutron star binaries is whether tidal
interaction can cause the individual stars to collapse into black holes during
inspiral.  To understand this issue better,
we study the dynamics of a cluster of collisionless
particles orbiting a non-rotating black hole, which is part of a widely
separated circular binary.  The companion body's electric- and magnetic-type
tidal fields distort the black hole and perturb the cluster, eventually
causing the cluster to collapse into the hole as the companion spirals in under
the influence of gravitational radiation reaction.  We find that magnetic-type
tidal forces do not significantly influence the evolution of the cluster as a
whole.  However, individual orbits can be strongly affected by these forces.
For example, some orbits are destabilized due to the addition of
magnetic-type tidal forces.  We find that
the most stable orbits are close to the companion's orbital plane and retrograde
with respect to the companion's orbit.
\end{abstract}

\section{Introduction}

Gravitational wave detectors will soon begin operation and may see signals from
sources such as binary neutron stars.  An important issue in the dynamics of these
binaries is whether tidal interaction between the neutron stars induces each one
to collapse to form a black hole during the late
stages of inspiral (\cite{mathews&wilson} and references therein).
Recent numerical simulations have shown that the stars either
have a very slight tendency towards individual collapse or are completely stable to such
collapse (see e.g.\ \cite{mathews&wilson,bgm,uryu&eriguchi}).
Given the complexity of relativistic hydrodynamic simulations, it is
useful to study simpler models in order to understand the physical underpinnings
of tidally induced collapse (or stability against collapse) in neutron star binaries.

One such model was considered by Shapiro and collaborators \cite{shapiro1,shapiro2}.
It consists of a cluster of
collisionless particles orbiting a non-rotating black hole, which is part
of a widely separated circular binary.
They found that as the hole's companion body spirals in slowly, its
tidal field induces inward collapse (into the hole) of the cluster.  However, they
restricted attention to the effect of the companion body's electric-type tidal field.
In the real physical situation of binary neutron stars, magnetic-type tidal fields
may play an important role in the stability of the stars \cite{favata&thorne}.
The purpose of this paper
is to investigate the effect of the companion body's magnetic-type tidal field on
the evolution of the particle cluster.

\section{Framework}

Consider a binary system composed of a non-rotating black hole of mass $m_1$
and a companion body
of mass $m_2$.  Let $m=m_1+m_2$ and $\mu=m_1 m_2/m$.  Denote the binary's orbital
separation by $b$.  Define the following Newtonian quantities:
the relative velocity $v=(m/b)^{1/2}$, and the orbital angular velocity 
\begin{equation}
\omega_N=(m/b^3)^{1/2}.
\label{omegaN}
\end{equation}
By assumption, $m\ll b$, so $v\ll 1$.

Assume the binary
is in a circular orbit that decays due to emission of Newtonian quadrupole
gravitational radiation.  This means the separation $b$ evolves as
\cite{peters,mtw}
\begin{equation}
b(t) = b_0(1-t/\tau_0)^{1/4},
\label{inspiral}
\end{equation}
where $\tau_0=(5/256)b_0^4(\mu m^2)^{-1}$.

Since the binary is widely separated,
the region near the black hole can be described by a tidally perturbed
Schwarzschild metric.  Cover this region by Schwarzschild coordinates
$(t,r,\theta,\phi)$.  Additional coordinate systems that will be useful are isotropic
coordinates $(t,x,y,z)$ and spherical isotropic coordinates $(t,R,\theta,\phi)$,
which are related to Schwarzschild coordinates via
\begin{equation}
r=R(1+m_1/2R)^2,\quad x=R\sin\theta\cos\phi,\quad y=R\sin\theta\sin\phi,
	\quad z=R\cos\theta.
\label{transform}
\end{equation}

Place a set of test particles at random positions on the sphere
$r=r_c\ll b$ and let their initial velocities
be in random directions tangent to the sphere.
In the test particle approximation,
each particle moves along a geodesic of the tidally perturbed Schwarzschild metric.
Set the magnitude of each particle's
initial velocity to yield a circular orbit when $m_2=0$.
Determine the cluster's time evolution by integrating the geodesic equation and
assuming the particles do not collide.

\section{Estimate of magnetic-type tidal force}

Given a 3+1 split of spacetime or a local reference frame, one can separate
the ten independent components of the Weyl tensor into an electric- and a
magnetic-type tidal tensor, each having five independent components (see
e.g.\ Secs.~V.A.2-3 in \cite{membrane}).
We are interested in the companion body's tidal fields as seen by an
inertial observer in the black
hole's local asymptotic rest frame \cite{thorne&hartle}.
This observer sees an electric-type tidal field whose typical
components have magnitude
${\cal E}\sim m_2/b^3$, where ``$\sim$'' means ``is of the order of''.
If the companion body was stationary with respect to the black hole, the observer
would not see a magnetic-type tidal field.  The motion of the companion at velocity
$v$ with respect to the black hole induces, via a Lorentz boost with velocity $v$,
a magnetic-type tidal field whose typical
components have magnitude ${\cal B}\sim v{\cal E}$
as seen by the inertial observer \cite{alvi}.

The magnitude of the magnetic-type tidal force experienced by
a particle moving in the local asymptotic rest frame is of order
$m_p v_p r{\cal B}\sim m_p v_p v r{\cal E}$ \cite{ni&zimmermann},
where $m_p$ is the particle's mass, $v_p$ is the particle's velocity
as measured by the inertial observer, and $r$ is the particle's orbital radius.
This is smaller by a factor of $v_p v$ than the electric-type tidal force
$m_p r{\cal E}$.

Now consider a particle cluster at $r=r_c$.  The representative particle velocity
of this cluster is $v_p=(m_1/r_c)^{1/2}$.  Shapiro and
collaborators set $r_c=6.9m_1$ and $m_1=m_2$,
and observe cluster collapse when the companion body reaches
$b\approx 15 m$ \cite{shapiro2}.
At that point, $v_p v=(m_1 m/r_c b)^{1/2}\approx 0.1$, so the magnetic-type tidal
forces on the particles are a significant fraction of the electric ones.
Furthermore, the two types of tidal forces typically point in different directions.
Therefore, the magnetic-type tidal field could have an
important effect on the dynamics of the cluster.  In the next sections, we
investigate this issue.

\section{Tidal perturbation}

The leading order electric- and magnetic-type tidal deformations of a non-rotating
black hole in a widely separated circular binary have been
calculated in \cite{alvi} using black hole perturbation theory.
The tidally perturbed metric
near the black hole is given in Schwarzschild coordinates $(t,r,\theta,\phi)$ by
(see Sec.~III in \cite{alvi})
\begin{eqnarray}
ds^2 &=& -\left(1-{2m_1\over r}\right)dt^2+\left(1-{2m_1\over r}\right)^{-1}dr^2
	+r^2(d\theta^2+\sin^2\theta d\phi^2)\nn\\
& & \mbox{}-{4m_2\over b^3}\sqrt{{m\over b}}\left(1-{2m_1\over r}\right)r^3
	dt\bigl[\cos\theta\sin(\phi-\omega t)d\theta
	+\sin\theta\cos(2\theta)\cos(\phi-\omega t)d\phi\bigr]\nn\\
& & \mbox{}+{m_2 r^2\over b^3}\left[3\sin^2\theta\cos^2(\phi-\omega t)-1\right]
	\Biggl[\left(1-{2m_1\over r}\right)^2 dt^2+dr^2\nn\\
& & \mbox{}+(r^2-2m_1^2)(d\theta^2+\sin^2\theta d\phi^2)\Biggr],
\label{schwmetric}
\end{eqnarray}
where
\begin{equation}
\omega=\omega_N\left(1-{\mu\over b}\right).
\label{omega}
\end{equation}
The metric perturbation has been written in
Regge-Wheeler gauge \cite{regge&wheeler} and the orbital plane of the companion has
been taken, without loss of generality, to be the black hole's equatorial plane.
The rotation rate $\omega$ of the
companion body's tidal field, as measured by an inertial observer in the black
hole's local asymptotic rest frame, differs from the Newtonian orbital angular
velocity $\omega_N$ because of precession effects \cite{alvi}.

The magnetic-type tidal perturbation on the black hole is embodied in the metric
components $g_{t\theta}$ and $g_{t\phi}$ in (\ref{schwmetric}).  Setting
$g_{t\theta}=0=g_{t\phi}$ yields the metric utilized in \cite{shapiro2}; this
metric contains only the electric-type tidal perturbation.  In what follows,
we sometimes switch off the magnetic-type tidal perturbation by setting
$g_{t\theta}=0=g_{t\phi}$, in order to isolate magnetic-type tidal effects
and to compare with the results in \cite{shapiro2}.

If we ignore radiation reaction, the binary's orbit remains circular and $\omega$
is constant.  In this situation, the vector field
$\vec{K}={\partial\over\partial t}+\omega{\partial\over\partial\phi}$
is a Killing field of the metric~(\ref{schwmetric}), and $\vec{K}\cdot\vec{u}$
is constant along each particle's geodesic worldline, where $\vec{u}$ is the
particle's four-velocity.  This constancy serves as an important check on our
numerics.  Indeed, we first turned off radiation reaction and verified
$\vec{K}\cdot\vec{u}$ was constant to sufficient accuracy along particle orbits.

In our actual simulations, we include the effects of radiation reaction on the
binary by substituting $b(t)$ from Eq.~(\ref{inspiral}) into~(\ref{schwmetric})
and (\ref{omega}).  We also replace $\omega t$ in~(\ref{schwmetric}) by
$\psi(t)=\int_0^t \omega(t')\, dt'$, where $\omega(t)$ is given by~(\ref{omega}),
(\ref{omegaN}), and~(\ref{inspiral}).
Furthermore, we transform~(\ref{schwmetric}) to isotropic coordinates $(t,x,y,z)$
in order to avoid difficulties with spherical coordinates.  This yields
\begin{eqnarray}
g_{tt} &=& -\left(\frac{1-m_1/2R}{1+m_1/2R}\right)^2
	+{m_2\over b^3}\left(1-{m_1\over 2R}\right)^4
	\left[3(x\cos\psi+y\sin\psi)^2-R^2\right],\nn\\
g_{tx} &=& {2m_2\over b^3}\sqrt{{m\over b}} \left
	(1-{m_1\over 2R}\right)^2 \left(1+{m_1\over 2R}
	\right)^4\left[(z^2-y^2)\sin\psi - xy\cos\psi\right],\nn\\
g_{ty} &=& {2m_2\over b^3}\sqrt{{m\over b}} \left
	(1-{m_1\over 2R}\right)^2 \left(1+{m_1\over 2R}
	\right)^4\left[(x^2-z^2)\cos\psi + xy\sin\psi\right],\nn\\
g_{tz} &=& {2m_2\over b^3}\sqrt{{m\over b}} \left
	(1-{m_1\over 2R}\right)^2 \left(1+{m_1\over 2R}
	\right)^4 (y\cos\psi- x\sin\psi)z,\nn\\
g_{ij} &=& \left(1+{m_1\over 2R}\right)^4 \Biggl(\delta_{ij}
	+{m_2\over b^3}\left[3(x\cos\psi+y\sin\psi)^2-R^2\right] \nn\\
& & \times\left\{\left[\left(1+{m_1\over 2R}\right)^4-{2m_1^2\over R^2}\right]
	\delta_{ij} -{2m_1\over R}\left(1+{m_1^2\over4R^2}
	\right){x^i x^j\over R^2}\right\}\Biggr),
\label{isotropicmetric}
\end{eqnarray}
where $R=(x^2+y^2+z^2)^{1/2}$.

The particles in the cluster follow geodesics of this metric.
We numerically integrate the geodesic equation
\begin{equation}
\ddot{x}^{\alpha} + {\Gamma^{\alpha}}_{\sigma\lambda}
	\dot{x}^{\sigma}\dot{x}^{\lambda} = 0
\label{geodesiceqn}
\end{equation}
to obtain the cluster's time evolution.  Here, an overdot represents a
derivative with respect to proper time.

\section{Numerical implementation and results}

We integrate Eq.~(\ref{geodesiceqn}) using a fourth order Runge-Kutta method
with adaptive stepsize control \cite{numrec}.  At $t=0$, we place the companion
at $b=60 m_1$ and put 1000 particles at Schwarzschild radius $r_c=6.9 m_1$.
We also set $m_1=m_2$.  In computing the Christoffel symbols
${\Gamma^{\alpha}}_{\sigma\lambda}$,
we ignore time derivatives of $b(t)$.  This is consistent with the treatment
in \cite{shapiro2} and is justified by the wide-separation/slow-inspiral
assumption.

We followed the cluster's evolution until $b=15m_1$ ($t=126068m_1$).
To check the accuracy of our code, we turned off radiation reaction (i.e.\ set
$b$ to be a constant) and verified that
$\vec{u}\cdot\vec{u}$ and $\vec{K}\cdot\vec{u}$ were conserved along particle
trajectories to a fractional accuracy of $10^{-7}$ during the entire evolution 
(i.e.\ from $t=0$ to $t=126068m_1$).

\begin{figure}
\begin{center}
\scalebox{.5}{\rotatebox{-90}{\includegraphics{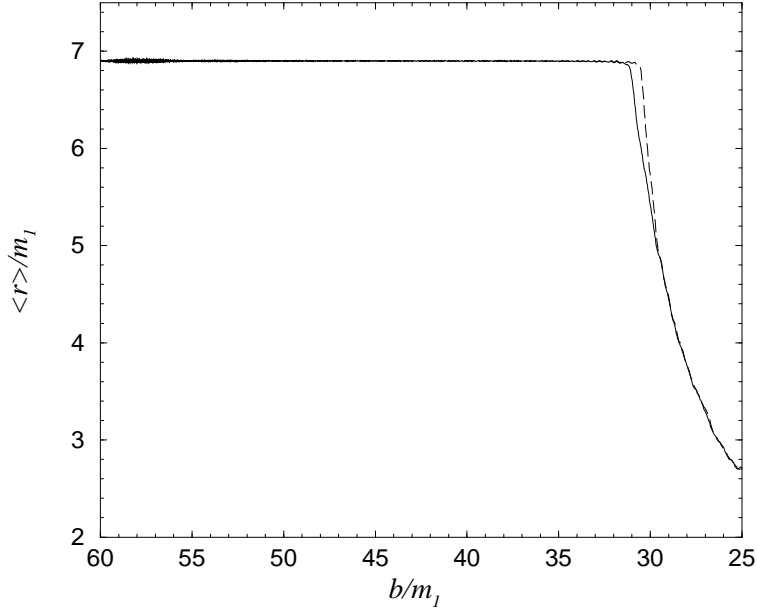}}}
\caption{\label{collapse1}
Cluster's average radius $\langle r\rangle$ as a function of binary separation
$b$.  The solid line represents evolution with both electric- and magnetic-type
tidal perturbations present, and the dashed line with only electric-type present.}
\end{center}
\end{figure}

The results of the simulation are presented in Fig.~\ref{collapse1}.
The cluster's average radius $\langle r\rangle$ is plotted against the binary's
separation $b$ for two cases: first, the magnetic-type tidal field and its
distortion of the black hole are ignored, i.e.\ the metric components $g_{ti}$
in (\ref{isotropicmetric}) are set to zero
(dashed line); and second, both electric- and magnetic-type tidal fields are
included, i.e.\ the full metric~(\ref{isotropicmetric}) is used (solid line).
The dashed line reproduces the corresponding result in \cite{shapiro2}.
The dynamics of the cluster as a whole is quite
similar in the two cases, even though individual orbits can be affected
significantly by the magnetic-type tidal perturbation (see below).
A large number of particles plunge into
the black hole at $b\approx 30m_1$, causing a sudden decrease of 
$\langle r\rangle$ (as seen in the figure).

It was observed in \cite{shapiro2} that particles initially orbiting
close to the equatorial
plane (which is also the companion's orbital plane) are more stable against collapse
into the black hole than those on initially polar orbits. 
Here we verify this observation in the presence of magnetic-type tidal forces.
For each particle that plunges into the black hole by the end of our simulation,
let $t_{{\rm plunge}}$ denote
the time at which the particle's orbital radius is $0.51 m_1$ ($2.000196 m_1$)
in isotropic (Schwarzschild) coordinates, and let 
$b_{{\rm plunge}}=b(t_{{\rm plunge}})$.
For each particle, let $\iota$ denote its
initial orbital inclination angle with respect to the equatorial plane.
This quantity is defined by
\begin{equation}
\cos\iota = \left.R^{-1}(x\dot{y}-y\dot{x})(\dot{x}^2+\dot{y}^2+\dot{z}^2)^{-1/2}
	\right|_{t=0}.
\end{equation}
On an unperturbed Schwarzschild background, this definition reduces to
$\cos\iota=L_z/L$, where $L_z$ is the $z$-component of the particle's
angular momentum and $L$ is its total angular momentum (both quantities are
conserved along particle trajectories).

\begin{figure}
\begin{center}
\scalebox{.5}{\rotatebox{-90}{\includegraphics{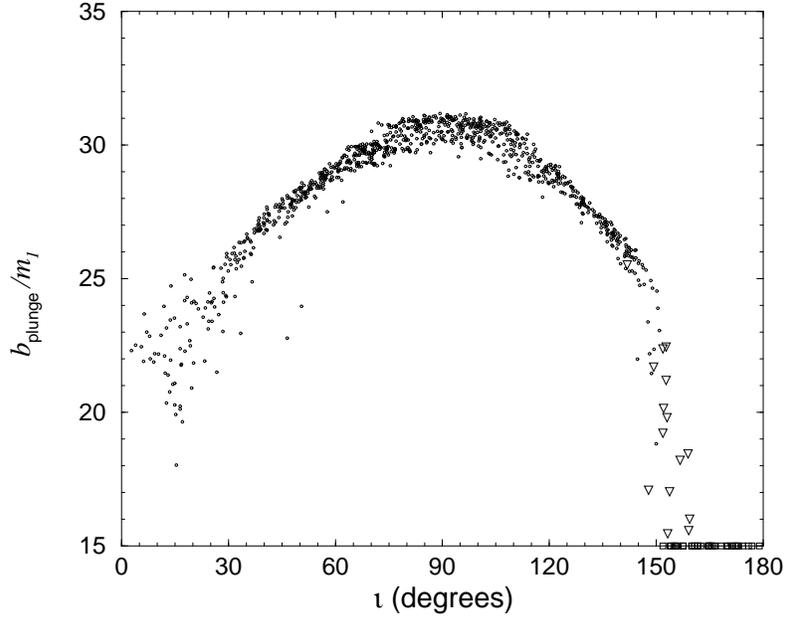}}}
\caption{\label{inclination}
Binary separation at plunge, $b_{{\rm plunge}}$, versus initial orbital inclination angle
$\iota$ for each particle.  Both electric- and magnetic-type tidal perturbations are
included.  Squares are particles that do not plunge; they are placed on the
$x$-axis.  Triangles are particles that are stable in the absence of the
magnetic-type tidal perturbation.}
\end{center}
\end{figure}

In Fig.~\ref{inclination}, we plot $b_{{\rm plunge}}$ versus $\iota$ for each particle
in the cluster.  The full metric~(\ref{isotropicmetric}) is used to compute the
cluster's evolution.  Particles that have not plunged by the end of our simulation are
represented by squares and are placed on the $x$-axis.  Particles that plunge in
the presence of the magnetic-type tidal perturbation but do not plunge in its
absence are represented by triangles.  There are no particles that plunge in
the absence of the magnetic-type tidal perturbation but do not plunge in its
presence.  The rest of the particles are represented by small circles.

A general trend in Fig.~\ref{inclination} is that the smaller the value of
$\sin\iota$, the more stable the orbit.
Orbits with $60^{\circ} \lesssim \iota \lesssim 120^{\circ}$
(nearly polar) are the least stable, while orbits with $\iota\grsim 165^{\circ}$
(retrograde with respect to the companion's orbit and nearly equatorial)
are the most stable.  Note that magnetic-type
tidal forces induce collapse in some particles that were stable in the absence of
these forces.

\begin{figure}
\begin{center}
\scalebox{.5}{\rotatebox{-90}{\includegraphics{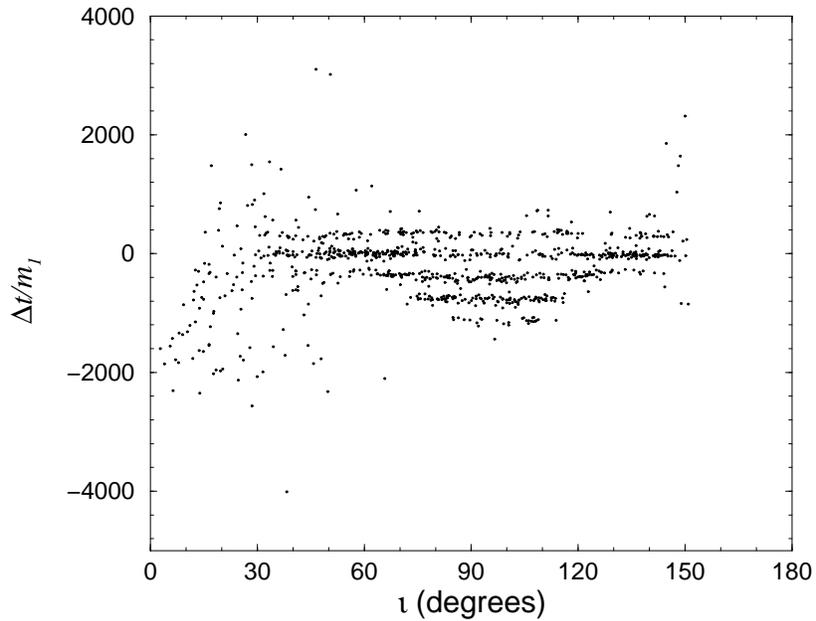}}}
\caption{\label{deltat}
Difference $\Delta t$ in plunge times with and without magnetic-type tidal
forces, plotted against initial orbital inclination angle $\iota$.
Positive values of $\Delta t$ correspond to a delayed plunge
due to magnetic-type tidal forces.}
\end{center}
\end{figure}

In order to study the effect of magnetic-type tidal forces on individual orbits,
we computed the time difference $\Delta t=t_{{\rm plunge}}^{em}-t_{{\rm plunge}}^e$,
where $t_{{\rm plunge}}^{em}$ ($t_{{\rm plunge}}^e$) is the plunge time with (without)
the magnetic-type tidal perturbation on the black hole, for particles that plunge
in both cases by the end of our simulation.  This time difference is
plotted as a function of the inclination angle $\iota$ in Fig.~\ref{deltat}.
It is clear from the figure that the magnetic-type tidal force can have a
significant influence on individual orbits.

\begin{figure}
\begin{center}
\scalebox{.5}{\rotatebox{-90}{\includegraphics{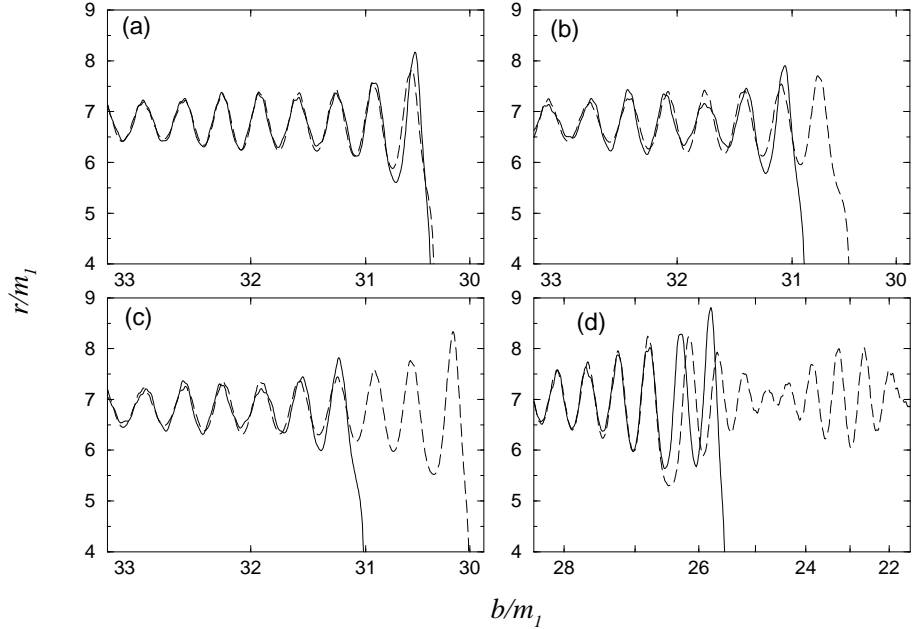}}}
\caption{\label{radial}
Radial oscillations before plunge for four selected particles:
(a) $\iota=81.4^{\circ}$, (b) $\iota=89.9^{\circ}$, (c) $\iota=89.6^{\circ}$,
and (d) $\iota=142^{\circ}$.  The vertical axis
is the Schwarzschild radius of the particle and the solid (dashed) lines represent
motion with (without) magnetic-type tidal forces.}
\end{center}
\end{figure}

For $50^{\circ} \lesssim \iota \lesssim 140^{\circ}$, the values of $\Delta t$
are roughly integral multiples of
\begin{equation}
T_r = 2 \pi r\left({r\over m_1}\right)^{1/2}\left(1-{6m_1\over r}\right)^{-1/2},
\label{Tr}
\end{equation}
which is the radial oscillation period of a nearly circular orbit outside $r=6m_1$
(see also Fig.~\ref{radial}).  
For $\iota\lesssim 30^{\circ}$, the plunge occurs at smaller values of $b$
(see Fig.~\ref{inclination}) and hence the particles experience 
stronger tidal forces.  As a result, the amplitudes of the particles' 
radial oscillations 
are large and the formula~(\ref{Tr}) for nearly circular orbits is no longer valid. 
In Fig.~\ref{deltat}, there are no points with $\iota \grsim 150^{\circ}$ because 
no particles with $\iota \grsim 150^{\circ}$ plunge in the absence of the magnetic-type
tidal perturbation (see Fig.~\ref{inclination}).

The last few radial oscillation periods before plunge for four selected particles
are shown in Fig.~\ref{radial}.  The vertical axis is the Schwarzschild radius of
the particle.  The solid line represents motion in the presence
of both electric- and magnetic-type tidal perturbations, and the dashed line in the
presence of only electric.  Cases (a), (b), and (c) correspond to particles 
with $\Delta t/T_r \approx 0$, -1, and -3 respectively.  Case (d) is an example of a 
particle that is stable (unstable) in the absence (presence) of the magnetic-type 
tidal perturbation.

\begin{figure}
\begin{center}
\scalebox{.5}{\rotatebox{-90}{\includegraphics{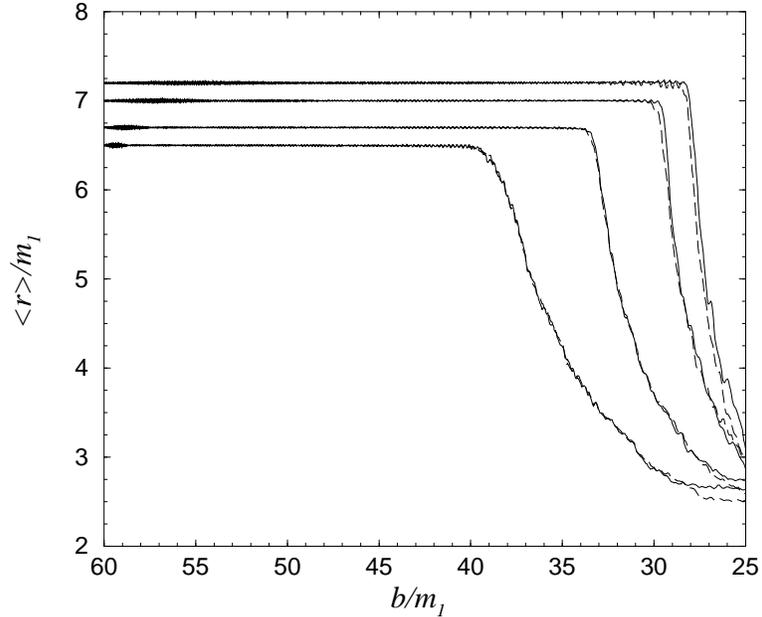}}}
\caption{\label{collapse4}
Cluster's average radius $\langle r\rangle$ as a function of binary separation
$b$ for $r_c=6.5$, 6.7, 7.0, and 7.2 (with 300 particles in the cluster).
The solid line represents evolution with both electric- and magnetic-type
tidal perturbations present, and the dashed line with only electric-type present.}
\end{center}
\end{figure}

We varied the initial cluster radius $r_c$
to investigate how this parameter affects the system's evolution.
Our results are displayed
in Fig.~\ref{collapse4}.  The cluster's average radius $\langle r\rangle$ is
plotted against binary separation $b$ for $r_c=6.5$, 6.7, 7.0, and 7.2.
We put 300 particles in the cluster for each of these simulations.  Once again,
solid (dashed) lines represent motion with (without) the magnetic-type tidal
perturbation.  Cluster collapse occurs in all cases and is qualitatively similar
with and without magnetic-type tidal forces.

Our simulations indicate that magnetic-type tidal forces do not significantly
influence the dynamics of the particle cluster as a whole, though individual orbits
can be strongly affected.

\section*{Acknowledgments}
We are grateful to Kip Thorne and Lior Burko for useful comments and conversations.
This research was supported in part by NSF grants PHY-0099568 and PHY-9796079,
and NASA grants NAG5-10707 and NAG5-4093.

\end{document}